\documentclass[aps,11pt]{revtex4}
\usepackage{dcolumn}
\usepackage{graphicx}
\usepackage{amsmath}
\usepackage{amsfonts}
\usepackage{amssymb}
\usepackage{psfrag}
\usepackage{wrapfig}
\usepackage{subfigure}
\usepackage{makeidx}
\usepackage{bm}
\usepackage{epsf}
\usepackage{hyperref}
\usepackage{color}
\usepackage{cases}
\makeatletter

\newcommand{\Rmnum}[1]{\expandafter\@slowromancap\romannumeral #1@}

\makeatother

\begin{document}

\title{Thermodynamics of  charged Lifshitz black holes with scalar hair}% Force line breaks with \\

\author{Shan Wu$^{1}$\footnote{18155514717@163.com}, Kai-Qiang Qian$^{1}$\footnote{kaiqiangqian@outlook.com}, Rui-Hong Yue$^{1}$\footnote{Corresponding author: rhyue@yzu.edu.cn}, Ming Zhang$^{2,3}$\footnote{Corresponding author: mingzhang0807@126.com} \\ and 
De-Cheng Zou$^{4}$\footnote{ dczou@jxnu.edu.cn}}

\affiliation{$^{1}$Center for Gravitation and Cosmology, College of Physical Science and Technology, Yangzhou University, Yangzhou 225009, China\\
$^{2}$Faculty of Science, Xi'an Aeronautical University, Xi'an 710077 China\\
$^{3}$School of Physics, Northwest University, Xi'an, 710069, China\\
$^{4}$College of Physics and Communication Electronics, Jiangxi Normal University, Nanchang 330022, China}

\begin{abstract}
In this work, we discuss the generalized Einstein-Maxwell-Dilaton gravity theory with a nonminimal coupling between the Maxwell field and scalar field. Considering different geometric properties of black hole horizon structure, the charged dilaton Lifshitz black hole solutions are presented in 4-dimensional spacetimes. Later, utilizing the Wald Formalism, we derive the thermodynamic first law of black hole and conserved quantities. According to the relationship between the heat capacity and the local stability of black hole, we study the stability of charged Lifshitz black holes and identify the thermodynamic stable region of black holes that meet the criteria.
\end{abstract}

\maketitle

\section{INTRODUCTION}
\label{sec:level1}

The Anti-de Sitter/Conformal Field Theory (AdS/CFT) correspondence~\cite{Maldacena:1997re,Lunin:2001jy,Maldacena:1998zhr,Witten:1998qj} provides a critical tool in theoretical physics to study strongly coupled relativistic field theories which states a gauge daulity  relating quantum field theory (QFT) and gravity. In more detail, the AdS/CFT correspondence corresponds  the classical dynamics of gravity to the quantum physics of strongly correlated systems. This principle which is also known as holographic duality or gauge/gravity correspondence is promoted to study nonrelativistic condensed matter theories~\cite{Hartnoll:2009sz}. As the gravity dual  to non-Lorentz invariant quantum field theories, the  Lifshitz spacetime \cite{Kachru:2008yh} is the natural setup. In the present work, the metric of Lifshitz spacetime is given by 
\begin{align*}
  \mathrm{d}s^{2} =- (\frac{r}{l})^{2 z} \mathrm{d} t^{2}+\frac{l^{2} }{ r^{2}}\mathrm{d} r^{2}+r^{2}\mathrm{d}^{2}\Omega.
\end{align*}
The metric is invariant under the scaling, therein, the anisotropy of the scaling symmetry is measured by the critical dynamical exponent $z$
\begin{align*}
  t\mapsto \lambda^{z}t,r\mapsto \lambda^{-1}r, x_{i}\mapsto \lambda x_{i}
\end{align*}

In general relativity(GR), the no-hair theorem has been an intriguing topic in black holes physics over decades. The first emergence of no-hair conjecture is mentioned in~\cite{Ruffini:1971bza}, which gives the types of matter fields allowed around regular black holes. 
To be more precise, Bekenstein puts forward a formal no-hair theorem~\cite{Bekenstein:1971hc}, which states that a black hole is completely described by the mass $M$, charge $Q$ and angular momentum $J$. However, the no-hair theorem has suffered some challenges, the hairy black hole solutions is allowed to appear in some gravity theories. In the colored black holes, the spherical solutions are supported by the Einstein-Yang-Mills SU(2) coupled system~\cite{Perry:1977wk,Galtsov:1989ip}.  
Black holes with dilaton hairs~\cite{Kanti:1995vq} and black holes with Skyrme hairs~\cite{Luckock:1986tr} have been found to serve as strong evidence against the no-hair theorem.  
During these years, the emerging phenomenon of spontaneous scalarization has attracted widespread attention. Spontaneous scalarization is caused by the instability of scalar fields, typically occurring in models with non-minimal coupling between a scalar field and  either additional (matter) fields. This phenomenon first appeared in research on the scalar-tensor model of neutron stars~\cite{Novak:1997hw}. Recently, Herdeiro \textit{et al.} proposed Einstein-Maxwell-scalar model with a specific nonminimal coupling between scalar field and Maxwell field~\cite{Herdeiro:2018wub}.  Later, the spontaneous scalarization through breaking scale invariance was worked out in Ref.~\cite{Herdeiro:2019yjy}, where the quantum corrections added in the form of a quartic term in the field strength breaks the natural Weyl invariance of Maxwell’s theory. Very recently, Alfredo \textit{et al.} \cite{Herrera-Aguilar:2020iti} and Moreira \textit{et al.} \cite{Moreira:2023zrl,Andrade:2024cdt} also presented some new exact black hole solutions in generalized Einstein-Maxwell-Dilaton setup, on account of breaking the spacetime isotropic scaling symmetry. There, these solutions are asymptotically Lifshitz for any dynamical critical exponent $z\geq1$. In the present work, we further explore this theory in the context of a charged Lifshitz background with different horizon structures for charged black holes.

On the other hand, black holes are generally considered thermodynamic systems with well-defined thermodynamic quantities, containing rich phase structures in spacetime and exhibiting critical phenomena similar to thermodynamic systems. Therefore, the study of black hole thermodynamics is a highly regarded topic. In 1973, the four laws of black hole mechanics were established~\cite{Bardeen:1973gs}, drawing an analogy between the area of the event horizon and surface gravity to entropy and temperature. Subsequently, various methods for calculating thermodynamic quantities emerged, such as the quasi-local conservation charge formula based on the Abbott-Deser-Tekin (ADT)~\cite{Abbott:1981ff} formalism covariantly extended by the Arnowit-Deser-Misner (ADM)~\cite{Arnowitt:1959ah} method , the counterterm method~\cite{Anabalon:2015xvl}, the Euclidean action method~\cite{Gibbons:1976ue}. Utilizing these theories, we can effectively demonstrate the first law of black hole mechanics. Wald first combined thermodynamic quantities with the first law of mechanics~\cite{Wald:1993nt,Gao:2001ut}, establishing the first law of black hole thermodynamics, where the mass of a black hole is determined by Wald's formalism. This law posits that the entropy of a black hole is the Noether charge~\cite{Wald:1993nt} , and the derivation of the entire process is generic, hence applicable to special theories such as supergravities and higher derivative gravities.
According to Wald's formalism, the essence of the first law of thermodynamics is the conservation law, which can be expressed as the superposition of contributions from different source fields, such as gravitational fields, vector fields, and scalar fields. 
In Ref.~\cite{Herrera-Aguilar:2020iti}, the idea of using the coupled term between Maxwell field and scalar field to induce charged Lifshitz background when looking for analytical Lifshitz black holes to discuss black hole thermodynamics. Here, we also explore analytical aspects of thermodynamics and its associated critical behavior of these scalarized lifshitz black holes.

This work is organized as follows. In Sec.~\ref{sec:level1}, we introduce the theoretical framework describing the nonminimal coupling between the scalar field and Maxwell field, and 
then present scalarized Lifshitz black holes with three different geometric properties. In Sec.~\ref{sec:level2},the conserved quantities and corresponding thermodynamics is provided, and we prove the first law of black hole thermodynamics by using the Wald formalism. Furthermore, we have also studied the thermodynamic stability of black holes. Finally, we end the paper with closing remarks in Sec.~\ref{sec:level3}. Throughout this text, we use Planck units, where  $G = c = \hbar  = 1$ in theoretical calculations, and utilize the $(-, +, +, +)$ signature for the metric.

\section{Solutions of charged lifshitz black holes}\label{sec:level1}

In this study, we will explore the solutions to a generalized Einstein-Maxwell-Dilaton  setup in four spacetime dimensions. Namely, the action of our framework reads
\begin{align} \label{lagrangian}
    S & = \frac{1}{16 \pi} \int \mathrm{d}^{4} x \sqrt{-g}\Big[ \frac{R-2\lambda}{2\kappa } -\frac{1}{2}\partial _{\mu} \phi\partial^{\mu }\phi-\frac{h\left ( \phi  \right ) }{4} F_{\mu \nu }F^{\mu \nu}\Big],
\end{align}
where $g = \mathrm{det}(g_{\mu \nu})$ denotes the metric determinant,{ }$R$ is the Ricci scalar,{ }$\lambda$ is the cosmological constant,{ }$\kappa$ is the Einstein's gravitational constant, { }$F_{\mu \nu}=\partial_{\mu}A_{\nu}-\partial_{\nu}A_{\mu}$ is the standard Maxwell field and we recall that  $h(\phi)$ is a nonminimal coupling funtion .

Varying the action { }(\ref{lagrangian}) with respect to the metric $g_{\mu\nu}$, dilaton field $\phi$, and electromagnetic potential $A_{\mu} $ leads to
\begin{align} 
    \nabla^2 \phi-\frac{1}{4}\frac{\mathrm{d} h}{\mathrm{d} \phi }F _{\mu\nu  } F^{\mu\nu } =0,\label{scalar-equ1}\\
    G_{\mu\nu}+\lambda g_{\mu\nu } -\kappa T_{\mu\nu}=0, \label{Einstein-equ1}\\
    \nabla_\nu (h(\phi) F^{\mu\nu})=0,\label{vector-equ1}
 \end{align}
where $G_{\mu\nu}=R_{\mu\nu}-\frac{1}{2} Rg_{\mu\nu}$ is the Einstein tensor, $T_{\mu\nu}$ is the energy-momentum tensor:
 \begin{eqnarray}
  T_{\mu\nu}=\partial_{\mu}\phi \partial _{\nu}\phi-\frac{1}{2}g_{\mu\nu}
  (\partial\phi  )^2+F_{\mu \alpha }F_{\nu}{ }^{\alpha }-\frac{1}{4} F_{\alpha \beta }{ }^{\alpha \beta } g_{\mu\nu}.
 \end{eqnarray}
 
Now we consider the following four-dimensional metric Ansatz
\begin{align}\label{ansatz}
    \mathrm{d}s^{2} & =- (\frac{r}{l})^{2z} f(r) \mathrm{d} t^{2}+\frac{l^{2} \mathrm{d}r^{2}}{r^{2}f(r) }+r^{2 }\mathrm{d}^{2} \Omega_k ,
\end{align}
where $z$ is dynamical critical exponent and $\mathrm{d}^{2} \Omega_k$ represents the line element given by
\begin{align}
\mathrm{d}^{2} \Omega_k & =\left\{
\begin{aligned} 
\mathrm{d}\theta^{2}+\sin^{2}\theta \mathrm{d}\phi^{2}  \quad    &\quad{k=1},\\
\mathrm{d}\theta^{2}+\mathrm{d}\phi^{2}      \quad  \qquad             &\quad{k=0},\\
\mathrm{d}\theta^{2}+\sinh^{2}\theta \mathrm{d}\phi^{2}     &\quad{k=-1}.\\
\end{aligned}
\right .
\end{align}
Note that the fields to inherit the spacetime isometries such that they are functions of the $r$ coordinate only. The vector potential is taken to be of the purely electrical form, thus, the field ansatzes are:
\begin{align}\label{Ar}
  \phi=\phi(r), \quad A=V(r)\mathrm{d}t.
\end{align}

Substituting the metric ansatz \eqref{ansatz} and Eq. \eqref{Ar} into the Maxwell equation (\ref{vector-equ1}), we  have
\begin{align} \label{vector-equ2}
  r{h}'(\phi){V}'(r){\phi }'(r)+h(\phi )\Big[-((-3+z){V}'(r)+r{V}''(r)\Big] =0.
\end{align}  
Under these requirements(\ref{Ar}), Eq.(\ref{vector-equ2}) becomes
\begin{align} \label{V First derivative}
{V}'(r)=\frac{Q}{h(\phi)} (\frac{l}{r} )^{3-z}, 
\end{align}
in which $Q$ is an arbitrary constant, and takes the role of the electric charge.

Dealing with the difference for components of Einstein field equations (\ref{Einstein-equ1}), one can find the identity
 \begin{eqnarray}
 && E^t{}_t- E^r{}_{r}=\frac{ f(r)(2-2z+r^{2} \kappa  {\phi}' (r)^2)}{l^2} =0,\label{Ett-Err}\\
 && E^r{}_{r}+ E^{\theta}{}_{\theta}=\frac{1}{2} (-\frac{2}{r^2} +4\lambda +\frac{2(2+3z+z^2)f(r)}{l^2} +\frac{r(5+3z){f}'(r)}{l^2}+\frac{r^2{f}''(r) }{l^2})=0.\label{Err+Etheta}
 \end{eqnarray} 
Then, we obtain the solution for the scalar field $\phi(r)$ in generalized Einstein-Maxwell-Dilaton gravity from Eq.\eqref{Ett-Err}
 \begin{eqnarray} 
 && \phi(r)=\phi_{0}+\sqrt{\frac{2(z-1)}{\kappa } } \ln(\frac{r}{l}),\label{phir}\\
 &&f(r)=\frac{kl^2}{r^2z^2} -\frac{2l^2\lambda }{2+3z+z^2} +\frac{a}{r^{ z+2}}+\frac{b}{r^{2(z+1)}}, \label{fr}
 \end{eqnarray}
 where $a$, $b$ and $\phi_{0}$ is an integration constant. Obviously, this solution is valid for $z \ge 1$. 

By inserting the expression of  the solutions $\phi(r) $, ${V}'(r)$ and $f(r)$ in the set of equations presented in the Einstein equation (\ref{Einstein-equ1}), as a consequence the nonminimal coupling funtion $h(r)$ becomes

\begin{eqnarray} \label{hr}
h(r)= -\frac{l^6Q^2r^{-2+2z}z(z+1)\kappa  }{2(bz^2(1+z)+l^2r^{2z}(-1+z)(-k(1+z)+r^2z\lambda ))},
 \end{eqnarray}
 Through the dilaton field equation, the relation of $r$ with respect to $\phi$ can be inversely solved as
\begin{eqnarray}
r(\phi )=e^{\frac{\sqrt{\frac{\kappa }{z-1} }(\phi -\phi_0)}{ \sqrt{2} } l },
\end{eqnarray}
 Plugging it into the nonminimal coupling function (\ref{hr}), the solution for $h[\phi (r)]$ can be represented like this
 \begin{align} \label{hphi}
  h[\phi(r)]=-\frac{\kappa z(z+1)Q^2l^6e^{-2\sqrt\frac{2\kappa }{z-1}(\phi -\phi _{0}  ) } }{2l^6(z-1)z\lambda -2kl^4(z^2-1)e^{-\sqrt\frac{2\kappa }{z-1}(\phi -\phi _{0} )}-2bl^{2-2z}z^2(z+1)e^{-\sqrt\frac{2\kappa }{z-1}(z+1)(\phi -\phi _{0} )}} ,
\end{align}

Suppose Maxwell equations accept trivial first integrals, the gauge potential is
\begin{align} \label{Vr}
 V(r)=V_{0}-\frac{2b}{Q\kappa l^{2z+3}} (\frac{l}{r} )^z+\frac{2k(z-1)}{Q\kappa z^2l} (\frac{r}{l})^z-\frac{2(z-1)\lambda l}{Q\kappa (z+1)(z+2) }(\frac{r}{l})^{z+2}  ,
\end{align}
where $V_{0}$ is an integration constant. The remaining  scalar field equation (\ref{scalar-equ1})are left for checking
 \begin{equation} \label{scalar-equ3}
  \frac{r^2\left ( \frac{r}{l}  \right )^{-2z}{h}'(\phi (r)){V}' (r)^2  }{2l^2} +\frac{r(r{f}'(r){\phi }'(r)+f(r)((3+z){\phi }'(r)+r{\phi }''(r)))}{l^2} =0.
 \end{equation}
 
 Altogether, in the present form, all solutions satisfy the whole set of equations of motion, and the metric function must satisfy the asymptotic condition
\begin{align} \label{lambda }
\lambda =-\frac{(z+2)(z+1)}{2l^2} 
\end{align}

\section{Thermodynamics of scalarized lifshitz black holes}\label{sec:level2}

In this section, we shall derive the thermodynamic quantities of charged Lifshitz black holes and apply Wald formalism to verify compliance with the first law of thermodynamics of black holes.Wald has showed the process of the variation of a Hamiltonian derived from a conserved Noether~\cite{Wald:1993nt}. Its application in the Einstein-Maxwell
theory and  the Einstein-scalar theory can be
found in Refs.~\cite{Gao:2003ys} and \cite{Gibbons:1996af}.
By analogy, we shall take this approach to verify the exactness of the conclusion in the Einstein-Maxwell-Dilaton theory by superimposing all the contributions. Following~\cite{Liu:2014tra}, we can first consider a general variation of the Lagrangian (\ref{lagrangian}):
\begin{equation}
\delta \mathcal{L}=\text { e.o.m. }+\sqrt{-g} \nabla_{\mu} J^{\mu},
\end{equation}

where e.o.m. denotes terms proportional to the equations of motion for the fields, and
\begin{equation}
J^{\mu}=g^{\mu \rho} g^{\nu \sigma}\left(\nabla_{\sigma} \delta g_{\nu \rho}-\nabla_{\rho} \delta g_{v \sigma}\right)
-\nabla^{\mu} \phi \delta \phi- h(\phi) F^{\mu \nu}\delta A_{\nu},
\end{equation}
\begin{align}
  Q_{(2)}^{\text {grav }} & =\frac{r^{z}(2zf(r)+rf'(r))}{l^{z+1}} r^{2} \Omega_{(2)} ,
\end{align}
\begin{align}
  Q_{(2)}^{\mathrm{A}} & =- h(\phi) V(r) V^{\prime}(r) (\frac{r}{l})^{1-z} r^{2} \Omega_{(2)},
\end{align}
\begin{eqnarray}
  i_{\xi} \Theta_{(3)}^{\text {grav }} =(\delta(\frac{r^{z}(2zf(r)+rf'(r))}{l^{z+1}})+\frac{2}{r} (\frac{r}{l})^{z+1} \delta (f(r))r^{2}\Omega_{(2)},
\end{eqnarray}
\begin{align}
  i_{\xi} \Theta_{(3)}^{\mathrm{A}} =- h(\phi) (\delta V(r)) V^{\prime}(r)(\frac{r}{l})^{1-z} r^{2} \Omega_{(2)} ,
\end{align}
\begin{align}
  i_{\xi} \Theta_{(3)}^{\phi}= f(r)(\frac{r}{l})^{z+1}\phi^{\prime} \delta \phi r^{2}\Omega_{(2)},
\end{align}
\begin{eqnarray} \label{firstlawintegral}
  \delta Q-i_{\xi} \Theta=-(h(\phi)V(r) \delta V'(r)(\frac{r}{l})^{1-z}+\frac{2}{r}(\frac{r}{l})^{z+1} \delta f(r)
  +f(r) (\frac{r}{l})^{z+1} \phi^{\prime}\delta\phi
)r^2\Omega_{(2)},
\end{eqnarray}
where $\xi=\partial_{t}$ is the  time-like Killing vector that is null on the horizon, $i_\xi$ denotes a contraction of $\xi^{\mu}$  on the first index of the $p-form$  $\Theta_{(p)}.$
Wald shows that the variation of the Hamiltonian with respect to the integration constants of a given solution is
\begin{align}
  \delta H=\frac{1}{16 \pi} \int_{\Sigma^{(2)}}\left(\delta Q-i_{\xi} \Theta\right),
\end{align}
where $\Sigma^{(2)}$ denotes two-boundaries of a Cauchy Surface,{ }one at infinity and one on the horizon $r_{h}$.

 The entropy of a black hole can be determined by the area law of entropy, which is applicable to nearly all types of black holes, including those that involve expansion within the framework of Einstein's theory of gravity. Thus, the solution of entropy is calculable 
\begin{eqnarray}
  S=\frac{\omega_{(2)}r_{h}^{2}}{4}.
\end{eqnarray}
Accordingly, the Hawking temperature reads
\begin{eqnarray}\label{TH}
  T_H=\frac{(\frac{r_{h}}{l})^{z+1}f'(r_{h})}{4\pi}.
\end{eqnarray}

The measurement choice is indispensable in order for the contribution of $\delta H_{r_{h}}$ on the horizon to be equivalent to $T_H \delta{S}$, which requires Maxwell field $A_{\mu}$ to vanish at the horizon.

Given the near-horizon form of the metric, we will have
\begin{align}
  f(r)=(r-r_{h})f^{\prime}(r_{h})+(r-r_h)^2{f}''(r_h) +\cdots,
\end{align}
where $r_{h}$ is the horizon radius.{ }Thus $\left.\delta f\right|_{r=r_{h}}=-\delta r_{h} f^{\prime}\left(r_{h}\right)$ and $ \delta H$  on the horizon is given by
 \begin{align}
  \delta \mathit{H} _{r_{h}}=\frac{ (\frac{r_h}{l})^{z+1}f'(r_h)\omega_{(2)}r_{h}\delta{r_h}}{8\pi},
 \end{align}
 since $T_H=(4 \pi)^{-1}(\frac{r_{h}}{l})^{z+1}f'(r_{h})$
 and $\delta{A}=4 \delta{S}=\delta({r_{h}}^{2}\omega_{(2)})=2 \omega_{(2)}r_{h}\delta r_{h}$ with $\omega_{(2)}=4 \pi$
 represents the volume of constant curvature subsurface described by $\mathrm{d}\Omega_{(2)}^{2}.$
 One can apply Gauss's law to calculate the charge of a black hole
\begin{equation}
  Q_{e}=\frac{1}{4 \pi}\int {h(\phi(r)) *F_{\mu \nu }n^{\mu } u^{\nu } r^2\mathrm{d}\Omega }=-\frac{ l^{2}Q}{16  \pi}  \omega _{(2)}.
\end{equation}
where $u$ and $n$ are timelike and spacelike normal unit vector given by
\begin{eqnarray}
&& n^{\mu }=\frac{1}{\sqrt{-g_{tt}}}  \mathrm{d}t=\frac{l^z}{r^z\sqrt{f(r)} } \mathrm{d}t\\
 && u^{\nu  }=\frac{1}{\sqrt{g_{rr}}}  \mathrm{d}r=\frac{r\sqrt{f(r)} }{l} \mathrm{d}r
\end{eqnarray}
The electrostatic potential difference between the horizon and infinity $(\Phi_e)$ associated with Maxwell field is defined as
 \begin{eqnarray}
 \Phi_{e}&=&V(\infty)-V(r_{h})=V(\infty) \nonumber\\
  &=&\lim_{r \to \infty} [V_{0}-\frac{2b}{Q\kappa l^{2z+3}} (\frac{l}{r} )^z+\frac{2k(z-1)}{Q\kappa z^2l} (\frac{r}{l})^z-\frac{2(z-1)\lambda l}{Q\kappa (z+1)(z+2) }(\frac{r}{l})^{z+2}] .
\end{eqnarray}

We then plug in the asymptotic condition $\lambda =-\frac{(z+2)(z+1)}{2l^2}$ and expand these functions $V(r)$, $f(r)$ ,$\phi(r)$, $h(r)$ in Eq.(\ref{firstlawintegral}) at infinity
\begin{eqnarray}\label{fr1}
  \ f(r)=1+\frac{kl^2}{r^2z^2}  +\frac{a}{r^{ z+2}}+\frac{b}{r^{2(z+1)}},
 \end{eqnarray}
 \begin{eqnarray}
  V(r)=V_0-\frac{2b}{Q\kappa l^{2z+3}} (\frac{l}{r} )^z+\frac{2k(z-1)}{Q\kappa z^2l} (\frac{r}{l})^z+\frac{(z-1) }{Q\kappa l }(\frac{r}{l})^{z+2} ,
 \end{eqnarray}
 \begin{align}
  \phi(r)= \sqrt[]{\frac{2(z-1)}{\kappa}}\ln_{}{(r/l)} +\phi_0 ,
 \end{align}
 \begin{align}
  h(r)=\frac{l^6Q^2z\kappa }{r^4(\frac{2l^2k(z-1)}{r^2}+z(z^2+z-2+\frac{2bz(r^ {-2z})}{r^2} ) )} .
 \end{align}
Variations  act on the parameter space including the integration constants, $V_0$ and $\phi_0$ fixed to make contributions from $V(r),\phi(r)$ vansih on the horizon, so the parameter space is solely spanned by $Q$, $a$ and $b$. For variation $f(r)$
 \begin{eqnarray}
  \delta{f}=\frac{\delta a}{r^{z+2} } +\frac{\delta b}{r^{2(z+1)} }  .
\end{eqnarray}
The rest of contributions are similar.
Substituting all these expansions and variations,and take the infinity limit and algebraic substitution $a=-m$
\begin{eqnarray}
  \delta H_{\infty}&=&\frac{1}{16 \pi} \int_{r \rightarrow \infty}\left(\delta Q-i_{\xi} \Theta\right)
 =\Big[\frac{\omega _{(2)}}{8 \pi l^{z+1}}\delta{m}+\frac{l^2\delta Q}{16\pi}\omega _{(2)}\lim_{r \to \infty}\Big[V_0-\frac{2b}{Q\kappa l^{2z+3}} (\frac{l}{r} )^z\nonumber\\ 
 &&+\frac{2k(z-1)}{Q\kappa z^2l} (\frac{r}{l})^z+\frac{(z-1) }{Q\kappa l }(\frac{r}{l})^{z+2}\Big]\nonumber\\
 &=&\delta M-\Phi_e\delta Q_e,
\end{eqnarray}
where $M=\omega_{(2)} \frac{m}{8 {\pi} l^{z+1}} $.

All other contributions vanish at infinity.
The first law of black hole thermodynamics arises from the equality between the variation of the Hamiltonian at asymptotic infinity and that on the horizon. i.e.
\begin{align}
  \delta \mathit{H} _{r_{h}}=\delta H_{\infty}.
\end{align}
 Indeed, substituting the solution into the Wald equation leads to the following first law of black hole thermodynamics 
\begin{eqnarray}
  \delta{M}=T \delta{S}+\Phi_e \delta Q_e.
\end{eqnarray}

In terms of horizon radius $r_h$, the mass $M$ of Wald formalism is obtained, according to the solution (\ref{fr1}) of $f(r)$,
\begin{align}
 m=\frac{b}{r_h^{z} } +\frac{kl^{2}r_h^{z}}{z^{2}} +r_h^{z+2}
\end{align}
then
\begin{align}
 M= \frac{\omega_{(2)}}{8 {\pi} l^{z+1}} (\frac{b}{r_h^{z} } +\frac{kl^{2}r_h^{z}}{z^{2}} +r_h^{z+2})
\end{align}
The Hawking temperature (\ref{TH}) of Lifshitz black holes can be calculated
\begin{align}
T_H=\frac{1}{4 \pi}(\frac{r_h}{l} )^{z+1} (-\frac{a(z+2)}{r_h^{z+3}}-\frac{2kl^2}{z^2 r_h^3 }-\frac{2b(1 + z)}{r_h^{2z+3}}  )
\end{align}

The local stability of a black hole depends on the sign of its heat capacity. A positive heat capacity corresponds to the black hole being locally stable under thermal fluctuations, whereas a negative heat capacity indicates the black hole is locally unstable. The heat capacity for a fixed charge is expressed as
\begin{align}
C_Q = \left ( \frac{\partial M }{\partial T_H}  \right ) _Q = \left ( \frac{\partial M }{\partial r_h}  \right ) _Q/\left ( \frac{\partial T_H }{\partial r_h}  \right ) _Q
\end{align}
where
\begin{eqnarray}
&&\left ( \frac{\partial M }{\partial r_h}  \right ) _Q = \frac{\omega_{(2)}}{8 {\pi} l^{z+1}} \left(-\frac{b z}{r_h^{z+1} } +\frac{kl^{2}r_h^{z-1}}{z} +(z+2)r_h^{z+1}\right),\\
&&\left ( \frac{\partial T_H }{\partial r_h}  \right ) _Q=\frac{1}{4\pi l^{z+1} } \Big[\frac{2a(z+2)}{r_h^3} - \frac{2k l^2  ( z-2)r_h^{z-3}}{z^2} + \frac{2b(z+2) (1 + z))}{r_h^{z+3}} \Big].
\end{eqnarray}

\begin{figure}[htb]
\centering
\subfigure[$0<r_h<3$]{\label{figure1} 
\includegraphics[width=2.8in]{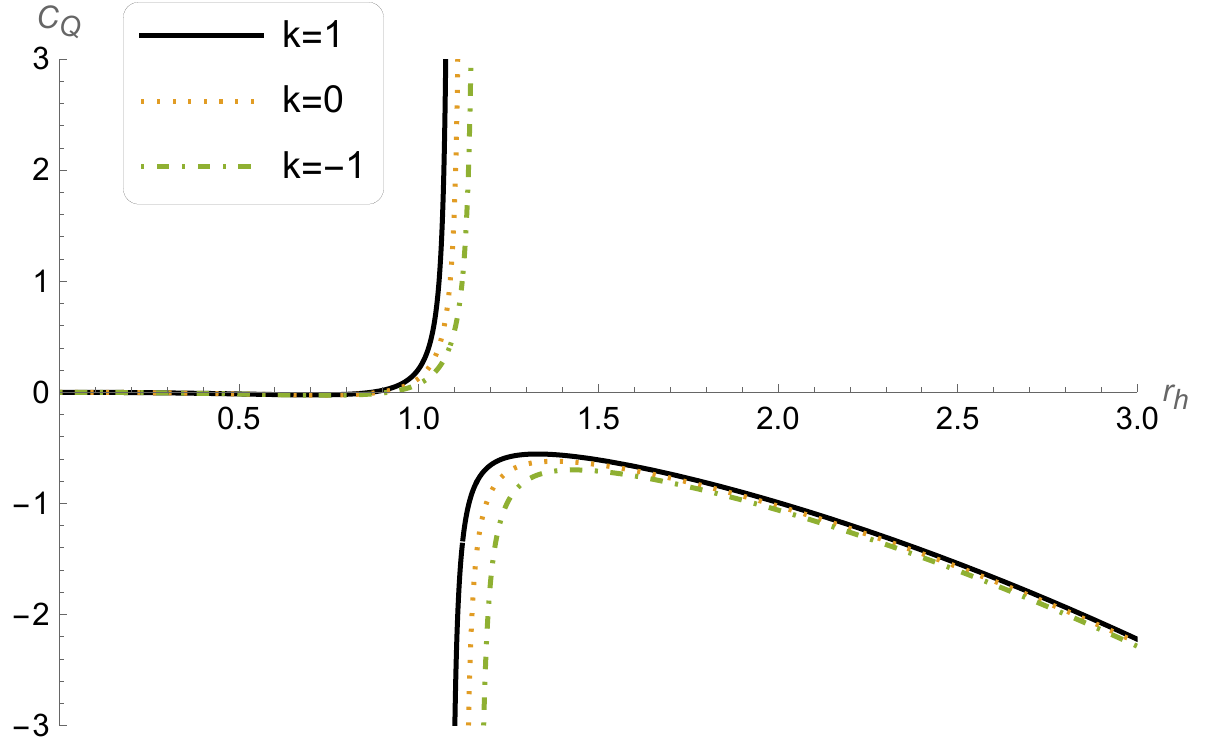}}
\hfill
\subfigure[$0<r_h<1$]{\label{figure2}
\includegraphics[width=2.8in]{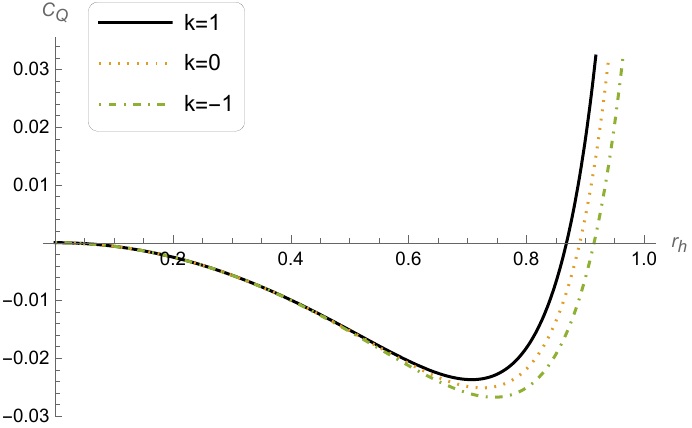}}
\hfill
\caption{Heat capacity $C_Q$  versus horizon radius $r_h$  for $b=1$, $\omega=1$, $l=1$, $z=2$}\label{figure3}
\end{figure}

\begin{figure}[htb]
\centering
\subfigure[$0<r_h<3$]{\label{figure5} 
\includegraphics[width=2.8in]{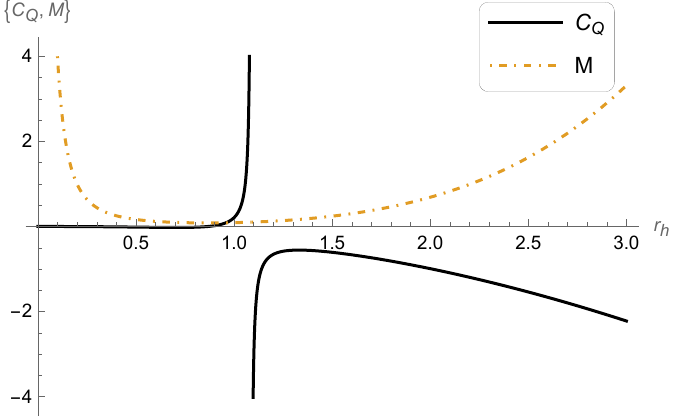}}
\hfill
\subfigure[$0<r_h<1$]{\label{figure6}
\includegraphics[width=2.8in]{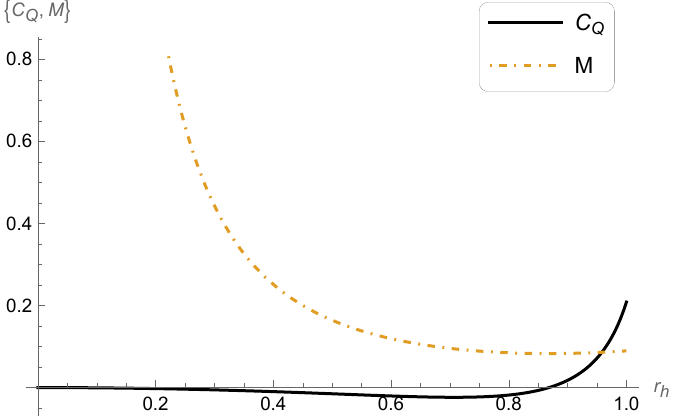}}
\hfill
\caption{Heat capacity $C_Q$ and $M$ versus horizon radius $r_h$  for $b=1$, $\omega=1$, $l=1$, $z=2$, $k=1$}\label{figure7}
\end{figure}

In Fig.~\ref{figure1}, we have plotted $C_Q$ as a function of horizon radius $r_h$ in the case of $k=1$, $k=0$, $k=-1$, the clearer image is displayed in Fig.~\ref{figure2}. Consider the heat capacity at three different geometric properties, their geometric characteristics are similar, we choose condition $k=1$ to illustrate. The graph of $M$ and $C_Q$ is plotted in Fig.~\ref{figure7}. By finding the extremum of $M$, we discover that $r_h$ reaches its minimum value at $r_{ext}\approx 0.868$, then steadily increases, always remaining a positive value. From Fig.~\ref{figure7}, the heat capacity $C_Q$ starts from zero at $r_{ext}\approx 0.868$ and increases sharply reaches a local maximum at  $r_m\approx 1.087$,  Therefore, the $C_Q$ maintains positive in the region $r_{ext}$ $<$$r_h$ $<$ $r_m$, and the black holes are locally stable in the domain $r_{ext}$ $<$ $r_h$ $<$ $r_m$.

\section{Closing Remarks}
\label{sec:level3}

In this paper, we have constructed static  asymptotically Lifshitz black hole solutions of Einstein–Maxwell–Dilaton theory in the presence of cosmological constant . In these theories, the presence of a scalar field is the reason for breaking the isotropy of space and time, and it is represented by $z$ to indicate the degree of departure from isotropy ($z>1$). The coupling relationship between the Maxwell field and the scalar field is expressed using a non-minimal coupling function $h[\phi(r)]$. Under the Lifshitz metric ansatz, the field equations have been derived from the variation principle. To solve these equations, we focus on the case where there are only charges in the vector field and assume that the scalar field has radial dependence only. We finally obtained analytical black hole solutions in the 4-dimensional. By applying Wald's formalism, we further examined the first law of thermodynamics of these black holes and obtained a thermodynamic conservation quantity. Studying the thermodynamics of these charged black hole solutions is of significant importance for the physical interpretation of their associated constants.
Subsequently, We summarize the stability issues for scalarized black hole solution. We then explored the stability of the Lifshitz black hole solution. Initially, we calculated the expressions of the mass $M$ and the heat capacity $C_Q$ of the system with respect to the event horizon radius $r_h$, and plotted their graphs. Based on this, we identified the stable region where the heat capacity is positive.

Furthermore, this work can be expanded and applied from various perspectives. We may explore other coupled theories such as the Einstein-Born-Infeld-scalar gravity within the Lifshitz background. Also, we can consider  the case where there is  magnetic vector field , this limitation of purely electrical form can be eased. Moreover, We believe that the Einstein–Maxwell–Dilaton theory makes a significant contribution to the study of spontaneous scalarization.

\vspace{1cm}
{\bf Acknowledgments}
\vspace{1cm}

D. C. Z is supported by the Natural
Science Foundation of China (NSFC) (Grant No.12365009) and Natural Science Foundation of Jiangxi Province (Grant No. 20232BAB201039).
M. Z. is supported by Natural Science Basic Research Program of Shaanxi (Program No.2023-JC-QN-0053).

\end{document}